\documentclass[a4paper,11pt]{article}
\usepackage[utf8]{inputenc}
\usepackage[T1]{fontenc}
\usepackage[english]{babel}
\usepackage[width=160mm,height=247mm]{geometry}
\usepackage{amsmath, amsfonts, amssymb, bm}
\usepackage[pdftex]{graphicx}
\usepackage{lmodern, indentfirst, cite}
\usepackage[squaren]{SIunits}
\usepackage[textfont={footnotesize},labelfont={color=blue,bf,sf},labelsep=endash]{caption}
\usepackage[colorlinks=true,linkcolor=blue,citecolor=blue,urlcolor=blue]{hyperref}

\usepackage[affil-sl]{authblk}
\title{\textbf{\Large A class of non-classicality and non-Gaussianity of photon added
three-mode GHZ-type entangled coherent states}}
\author[1,2]{Larbi Jebli\thanks{\href{mailto:larbi.jebli@gmail.com}{larbi.jebli@gmail.com}}}
\author[1,2]{Rachid Houça\thanks{\href{mailto:r.houca@uiz.ac.ma}{r.houca@uiz.ac.ma}}}
\author[3,4]{Mohammed Daoud\thanks{\href{mailto:m\textunderscore{}daoud@hotmail.com}{m\textunderscore{}daoud@hotmail.com}}}
\affil[1]{Team of Theoretical Physics, Laboratory L.P.M.C., Department of Physics, Faculty of Sciences, Chouaib Doukkali University, PO Box 20, 24000 El Jadida, Morocco}
\affil[2]{Team of Theoretical Physics, Laboratory L.P.T.H.E., Department of Physics, Faculty of Sciences,
Ibn Zohr University, PO Box 8106, Agadir, Morocco}
\affil[3]{Department of Physics, Faculty of Sciences, University Ibn Tofail, Kenitra, Morocco}
\affil[4]{Abdus Salam International Centre for Theoretical Physics, Miramare, Trieste, Italy}
\date{}

\newcommand{\beq}{\begin{equation}}
\newcommand{\eeq}{\end{equation}}
\newcommand{\lb}{\label}
\begin{document}
\begin{titlepage}
	\newgeometry{width=175mm, height=247mm}
    \maketitle
    \thispagestyle{empty}
    \vspace{3cm}
 \begin{abstract}
 {\color{black}In this paper, We investigate three-mode photon-added Greenberger-Horne-Zeilinger (GHZ) entangled coherent states by repeatedly operating the photon-added operator on the GHZ entangled coherent states. The product of two Laguerre polynomials is demonstrated to be connected to the normalizing constant. The influence of the operation on the non-classical and non-Gaussian behavior of the GHZ entangled coherent states is investigated. Sub-Poissonian statistics, such as Mandel's parameter and the negativity of the Wigner function, show that non-classical properties can enhance GHZ entangled coherent states. Finally, the occurrence of the anti-bunching phenomena in this class of tripartite excited states is studied using the second-order correlation function.}
 \end{abstract}
\vspace{3cm}
\end{titlepage}
\section{Introduction}
In quantum information processing tasks, we usually encode information in multipartite quantum states with high amount of quantum correlations between the different parties of the system \cite{Hu2010,Zhang2012,Karimi2016,Hu2016}. In this sense, it is always needed to find the ways to enhance and to protect the degree of entanglement between the components of the quantum system employed to implement quantum protocols especially in presence of effects such as photon-substraction, photo addition, and their superposition, mixing local squeezing and local displacement. In this context, recently some useful tools to generate and to enhance the entanglement degree between two and three modes of quantized radiation field.
This is essentially due to the advent of experimental quantum optics of quantum optics allows to produce and manipulate various non-classical optical fields \cite{Braunstein03,Marek}. The photon addition and subtraction modeled mathematically by the actions of bosons creation operators $a^\dag$ and annihilation operators $a$, have prompted significant investigations \cite{Agarwal1991,Zavatta,Kitagawa}. As quantum entanglement is a crucial part of quantum information theory and is at the heart of many quantum technologies, a lot of attention was dedicated to the generation of this resource using optical states to encode the quantum information to be processed to implement quantum protocols such as \cite{NC-QIQC-2000,DiVincenzo,Bennett}, quantum teleportation \cite{Grover,Braunstein}, quantum dense coding \cite{Bennett92, Braunstein00}, quantum key distribution \cite{Ekert}, quantum cryptography \cite{Bennett92,Ekert}, and many other applications. In exploiting quantum correlations in multi-particle quantum system, a classification of the states according their degree of entanglement was considered in several works. The states violating Bell inequalities were studied in \cite{Bell64,Clauser69,Clauser74,Jeong}. The W-type and GHZ-type entangled states were considered in \cite{Greenberger,Dur}.

{\color{black}{In another development, much attention has been paid to the multi-mode coherent states, Gerry et al. \cite{Gerry26} propose a method for generating ECSs of a two-mode field. Moreover, ones have studied the optimal quantum information processing via GHZ-type \cite{Jeong27} and W-type ECSs \cite{Nguyen28}. Jeong's group \cite{Jeong27} present the generation of GHZ-type ECSs using beam splitters (BSs) with a single-mode CSS and demonstrate Bell inequality violations for GHZtype ECSs. Some theoretical schemes have been proposed to generate the GHZ-type ECS in cavity fields \cite{Song29,Yuan30}. Kuang's group \cite{Xu34,Zhou35} propose single-mode and two-mode excite ECSs and their generation via cavity QED. The optical generation of excited ECSs is also investigated by using the BSs and the type-I beta-barium borate (BBO) crystal \cite{Li36,Ren37}. Furthermore, the photon-added coherent state (PACS) was first proposed by Agarwal and Tara \cite{Agarwal1991}, they
presented a hybrid non-classical state called a PACs which exhibits an intermediate property between a classical coherent state (CS) and a purely quantum Fock state (FS) \cite{Agarwal1991}. More recently, by using a type-I BB) crystal, a single photon detector (SPD) and a balanced homodyne detector, Zavatta et al. experimentally created a single-photon-added coherent state (SPACS) which allowed them to first visualize the classical-to quantum transition process \cite{Zavatta9}. In the other hand, the authors \cite{O Safaeian} try to present a general formalism for the construction of deformed photon-added nonlinear coherent states (DPANCSs), which in special case lead to the well-known PACS. Recently, \cite{[2019]} introduced a new kind of photon-added entangled coherent states (PA-ECSs), by performing repeatedly an f-deformed photon-addition (DPA) operation, on each mode of the entangled coherent states (ECSs). By choosing a particular deformation function, as a result, they study how the entanglement properties can be enhanced by DPA operation. Furthermore, their findings indicate that there is a family of coherent amplitudes $\left \vert \alpha \right \vert $ for which the two-mode DPA operation preserves the maximal entanglement of the odd ECSs while the non-deformed photon-addition operation suppresses it. In the other study, Mojaveri et al. want to see how adding photons to two-qutrit entangled states affects them. They give a general study of non-classical features including photon statistics and entanglement for this purpose, with a focus on the control role of the shift parameter in these states \cite{Mojaveri14}.

The experimental generation of non-classical features of quantum states based on the superposition of photon-added even/odd coherent states has recently received a special interest. However, adding photons to any quantum state, such as the thermal state, coherent state, squeezed vacuum state, is in general a challenging task and can produce classical states. The features of the photon-added squeezed vacuum state were investigated in \cite{Zhang}, photon-added coherent states, and photon-added thermal states considered in several works \cite{Marek,Lee,Jones}. The formalism of photon added coherent states or excited coherent states were considered in \cite{Li-Yun,Nath,yuan2009}. Many authors have examined the effects of adding quanta to coherent states \cite{Agarwal1991}, of successive elementary one-photon excitations of a coherent state and modified photon-added coherent states \cite{Liang}.

This study aims to investigate the non-classical and non-Gaussian characteristics of photons that have been introduced to a multi-mode (GHZ)
entangled coherent state. The purpose is to distil the better one for quantum information processing implementation and compare it to a
single-mode excited GHZ-type entangled coherent state \cite{Li2009}, measuring quantum correlations in non-classical multipartite coherent states \cite{yuan2009,Li2009, Hu2009, Zhang2016}. In this paper, we propose A class of three-mode excited GHZ-type entangled coherent states (EGHZECSs) that are produced by conducting creation operator operations on GHZ-type ECSs and then decomposing them into the additional photon even (odd) coherent state.}}

Quantum non-classicality of single-mode and multi-mode systems is a fundamental property of quantum optics and quantum physics. The quasi-probability functions, which include the Glauber-Sudarshan $P$-function \cite{GlauberS,Sudarshan}, the Sudarshan $Q$-function (Husimi) \cite{Husimi}, and the Wigner function (WF) \cite{Wigner}, are essential tools in characterizing non-classicality through their negative values. The negativity of the Wigner function distribution is a powerful tool to characterize non-classicality.

This paper is organized as follows. In Section 2, we introduce three-mode GHZ entangled coherent states. We give the expression of the superpositions of multipartite entangled coherent states and we derive analytically the expression of three-mode photon-added GHZ entangled coherent states (PAGHZECSs), which are obtained by adding photons to a general multi-mode GHZ-coherent state. In Section 3, we investigate the non-classical photon statistical properties of multiple-photon-added entangled coherent states, the associated sub-Poissonian photon statistics, and the second-order correlation function. These quantifiers are all obtained by applying the partial negativity of the Wigner function for photon-added three-mode GHZ entangled coherent states. The obtained results are illustrated by numerical analysis. Finally, we end up with some closing remarks.

\subsection{\protect \bigskip Superpositions of multipartite entangled
coherent states}

Any multipartite state associated with a multi-partite bosonic modes system may be expressed as a superposition of entangled coherent states. Here, we consider $N$ mode GHZ coherent states defined by
\begin{equation}
\left \vert \psi _{\varphi }\right \rangle =\mathcal{N}_{\varphi }^{0}\left(
\left \vert \alpha \right \rangle _{1}\otimes \left \vert \alpha \right
\rangle_{2} \otimes ...\otimes \left \vert \alpha \right \rangle _{N}+\exp
\left( i\varphi \right) \left \vert -\alpha \right \rangle _{1}\otimes \left
\vert -\alpha \right \rangle_{2} \otimes ...\otimes \left \vert -\alpha \right
\rangle _{N}\right) ,\label{vect-0}
\end{equation}
where $\left \vert \alpha \right \rangle _{1}\otimes \left \vert \alpha
\right \rangle _{2} \otimes ...\otimes \left \vert \alpha \right \rangle
_{N}=\left \vert \alpha ,\alpha ,...,\alpha \right \rangle _{1,2,...,N}$ with $%
\left \vert \alpha \right \rangle _{i}=\exp \left( -\frac{\left \vert \alpha
\right \vert ^{2}}{2}\right) \sum \limits_{n_{i}=0}^{+\infty }\frac{\alpha^{n_{i}}}{\sqrt{n_{i}!}}\left \vert n_{i}\right \rangle $ represents a bosonic coherent state of amplitude $\alpha $ in $i-$th modes, and $\mathcal{N}$ denotes the normalization factor given by
\begin{equation}
\mathcal{N}_{\varphi }^{0}=\left \{ 2\left[ 1+\kappa ^{n}\cos \left( \varphi
\right) \right] \right \} ^{-1/2}  \label{Normaliz-n}
\end{equation}
where the quantity $\kappa $, represents the overlap between the states $\left \vert \alpha \right \rangle $ and $\left \vert -\alpha \right \rangle $. It is given by
\begin{equation*}
\kappa =\left \langle \alpha |-\alpha \right \rangle =\exp \left( -2\left
\vert \alpha \right \vert ^{2}\right)
\end{equation*}
In this work, we shall consider three mode coherent states $(N=3)$.
\subsubsection{Three-mode photon-added GHZ-type entangled coherent states}
The photon-added or excited GHZ entangled coherent states may be generated by repeated action of the creation operator on the three-mode GHZ-type ECSs. The resulting excited GHZ-type ECSs are given by
\begin{equation}
\left \vert \psi _{\varphi }^{r,s,t}\right \rangle =\left( \mathcal{N}%
_{\varphi }^{r,s,t}\right) ^{-1/2}\left( a_{1}^{\dagger r}a_{2}^{\dagger
s}a_{3}^{\dagger t}\right) \left( \left \vert \alpha \right \rangle
_{1}\otimes \left \vert \alpha \right \rangle _{2}\otimes \left \vert \alpha
\right \rangle _{3}+\exp \left( i\varphi \right) \left \vert -\alpha \right
\rangle _{1}\otimes \left \vert -\alpha \right \rangle _{2}\otimes \left
\vert -\alpha \right \rangle _{3}\right)  \label{vect-PA}
\end{equation}
where $\mathcal{N}_{\varphi }^{r,s,t}$ is the normalization factor. It can be computed from (\ref{vect-PA}) by employing \cite{Li2009,yuan2009}
\begin{equation}
\left \langle \alpha|a_{i}^{\dagger m}a_{i}^{m}|\alpha\right
\rangle =m!L_{m}\left( -\left \vert \alpha\right \vert ^{2}\right) ,%
\text{ \  \ }\left \langle \alpha|a_{i}^{\dagger m}a_{i}^{m}|-\alpha\right \rangle =m!\kappa L_{m}\left( \left \vert \alpha\right
\vert ^{2}\right) ,  \label{overlap}
\end{equation}
where $L_{m}\left( x\right) \equiv L_{m}^{0}\left( x\right)$ stands for the $m$-order Laguerre polynomial $L_{m}^{k}$ for $k=0$. The polynomial $L_{m}^{k}$ are defined by  \cite{Agarwal1991,duc2008}
\begin{equation}
L_{m}^{k}\left( x\right) =\sum_{l=0}^{m}\frac{\left( -1\right) ^{l}\left(
m+k\right) !x^{l}}{l!\left( l+k\right) !\left( m-l\right) !}
\label{Laguerre}
\end{equation}

We define the state $\vert \alpha,m \rangle_i$ as follows
\begin{equation}
\left \vert \alpha,m\right \rangle_{i} =\frac{\kappa}{\left[
m!L_{m}\left( -\left \vert \alpha\right \vert ^{2}\right) \right] ^{1/2}%
}\sum_{l=0}^{\infty }\frac{\alpha ^{l}\sqrt{l+m}}{l!}\left \vert l+m\right
\rangle_{i}  \label{Fockstates}
\end{equation}
The photon-added GHZ-type ECSs $\left\vert \psi _{\varphi }^{r,s,t}\right \rangle$ is a three-component entangled state between modes $1$, $2$ and $3$. The normalization factor in eq(\ref{vect-PA}) is given by
\begin{equation}
\mathcal{N}_{\varphi }^{r,s,t}=\left \{ 2r!s!t!\left[ p\left( -\alpha
,r\right) p\left( -\alpha,s\right) p\left( -\alpha
,t\right) +p\left( \alpha,r\right) p\left( \alpha,s\right) p\left( \alpha,t\right) \cos (\varphi )\right] \right
\}  \label{Normaliz PA}
\end{equation}
where \begin{equation}
p\left( -\alpha,m\right) =L_{m}\left( -\left \vert \alpha\right \vert ^{2}\right) \text{, \  \  \ }p\left( \alpha,m\right)=L_{m}\left( \left \vert \alpha\right \vert ^{2}\right) \kappa \end{equation}
In the particular case $t=0$, which corresponds to the case of two-mode entangled coherent states, the equation (\ref{Normaliz PA}) leads to
\beq\lb{ff}
\mathcal{N}_{\varphi
}^{r,s,0}=\left \{ 2r!s!\left[ \left( L_{r}\left( -\left \vert \alpha
\right \vert ^{2}\right) L_{s}\left( -\left \vert \alpha\right
\vert ^{2}\right) +\kappa^{3}L_{r}\left(
\left \vert \alpha\right \vert ^{2}\right) L_{s}\left( \left \vert
\alpha\right \vert ^{2}\right) \cos (\varphi )\right) \right] \right \}
\eeq
and for $s=t=0$, the equation \eqref{ff} takes the form
\beq
\mathcal{N}_{\varphi
}^{r,0,0}=\left \{ 2r!\left[ \left( L_{r}\left( -\left \vert \alpha
\right \vert ^{2}\right) +\kappa^{3}L_{r}\left(
\left \vert \alpha\right \vert ^{2}\right) \cos (\varphi )\right) %
\right] \right \}
\eeq
One can verify that for $r=s=t=0$, the normalized factor $\mathcal{N}_{\varphi }^{PA}$ of the three-mode GHZ ECSs reduces to $\mathcal{N}_{\varphi }^{0}=\left \{ 2\left[ 1+\kappa ^{3}\cos \left( \varphi \right)\right] \right \} ^{-1/2}$, and the GHZ entangled coherent states \eqref{vect-0} are recovered.

\section{Non-classical Photon statistical Properties}

In this section, we shall consider the photon addition effects on the nonclassical photon statistical features of the PAGHZECSs. The nonclassical properties of the PAGHZECSs will be studied in terms of the Wigner function negativity, Mandel's $Q$ parameter and the cross-correlation function.

\subsection{The Partial Negativity of the Wigner Function for the
photon-added Three-Mode GHZ entangled coherent states (PAGHZECSs)}
First, we determine at the Wigner function's negativity condition for three-mode entangled coherent states. In general, when the Wigner function exhibits partial negativity in the phase space, the quantum state is deem strongly nonclassical. The negativity of Wigner function's is a good indicator for displaying nonclassical features (and investigating non-classicality) of quantum states and describing decoherence of quantum states when exposed to the environment effects. The Wigner function associated wit a given density operator $\rho$ is defined by \cite{Li2009}.

\begin{eqnarray}
W\left( \eta ,\gamma ,\delta \right) &=&\exp \left[ 2\left( \left \vert \eta
\right \vert ^{2}+\left \vert \gamma \right \vert ^{2}+\left \vert \delta
\right \vert ^{2}\right) \right] \int \frac{d^{2}z_{1}}{\pi ^{2}}\frac{%
d^{2}z_{2}}{\pi ^{2}}\frac{d^{2}z_{3}}{\pi ^{2}}\left \langle
-z_{1},-z_{2},-z_{3}\right \vert \rho \left \vert z_{1},z_{2},z_{3}\right
\rangle  \label{Wigner F} \\
&&\times \exp \left[ 2\left( \eta z_{1}^{\ast }-\eta ^{\ast }z_{1}\right) %
\right] \exp \left[ 2\left( \gamma z_{2}^{\ast }-\gamma ^{\ast }z_{2}\right) %
\right] \exp \left[ 2\left( \delta z_{3}^{\ast }-\delta ^{\ast }z_{3}\right) %
\right]  \notag
\end{eqnarray}

\noindent The density operator of three-mode added coherent states in equation (\ref{Wigner F}) takes the form
\begin{eqnarray} \label{density OPA}
\rho &=&\left \vert \psi _{\varphi }^{r,s,t}\right \rangle \left \langle \psi_{\varphi }^{r,s,t}\right \vert \\ \nonumber
&=& \left( \mathcal{N}_{\varphi }^{r,s,t}\right) ^{2}\left( a_{1}^{\dagger
r}a_{2}^{\dagger s}a_{3}^{\dagger t}\right) \left( \left \vert \alpha
\right
\rangle _{1}\otimes \left \vert \alpha \right \rangle _{2}\otimes
\left
\vert \alpha \right \rangle _{3}
+
\exp \left( i\varphi \right)
\left
\vert -\alpha \right \rangle _{1}\otimes \left \vert -\alpha
\right
\rangle _{2}\otimes \left \vert -\alpha \right \rangle _{3}\right)\\ \nonumber
&\times&
\left( \left
\langle \alpha,\alpha,\alpha\right \vert +\exp
\left( -i\varphi \right) \left \langle -\alpha,-\alpha,-\alpha
\right
\vert \left( a_{1}^{r}a_{2}^{s}a_{3}^{t}\right) \right) \nonumber
\end{eqnarray}
It can be rewritten as
\begin{eqnarray}
  \rho &=& \left( \mathcal{N}_{\varphi }^{r,s,t}\right) ^{2}\left( a_{1}^{\dagger
r}a_{2}^{\dagger s}a_{3}^{\dagger t}\right) \left \vert \alpha,\alpha
,\alpha\right \rangle \left \langle \alpha,\alpha,\alpha
\right \vert \left( a_{1}^{r}a_{2}^{s}a_{3}^{t}\right) +\exp \left(
i\varphi \right) \left( a_{1}^{\dagger r}a_{2}^{\dagger s}a_{3}^{\dagger
t}\right) \\ \nonumber
&\times&
\left \vert -\alpha,-\alpha,-\alpha\right \rangle
\left \langle \alpha,\alpha,\alpha\right \vert \left(
a_{1}^{r}a_{2}^{s}a_{3}^{t}\right) \\ \nonumber
   &=& \left( \mathcal{N}_{\varphi }^{r,s,t}\right) ^{2}
   \Bigg[ \exp \left(
-i\varphi \right) \left( a_{1}^{\dagger r}a_{2}^{\dagger s}a_{3}^{\dagger
t}\right) \left \vert \alpha,\alpha,\alpha\right \rangle
\left \langle -\alpha,-\alpha,-\alpha\right \vert \left(
a_{1}^{r}a_{2}^{s}a_{3}^{t}\right)\\ \nonumber
&+&
 \left( a_{1}^{\dagger r}a_{2}^{\dagger
s}a_{3}^{\dagger t}\right)
 \left \vert -\alpha,-\alpha,-\alpha
 \right \rangle \left \langle -\alpha ,-\alpha ,-\alpha
\right \vert
\left( a_{1}^{r}a_{2}^{s}a_{3}^{t}\right) \Bigg]
\end{eqnarray}

\noindent where $\left \vert z_{1},z_{2},z_{3}\right \rangle =\left \vert
z_{1}\right \rangle \otimes \left \vert z_{2}\right \rangle \otimes \left \vert z_{3}\right \rangle $ is the
three-mode coherent state. Using the overlap between two coherent states $\left \vert
\alpha\right \rangle$ and $\left \vert \beta\right \rangle $

\beq
\left
\langle \alpha |\beta \right \rangle =\exp
\left[ -\frac{\left \vert \alpha \right \vert ^{2}}{2}-\frac{\left \vert
\beta \right \vert ^{2}}{2}+\alpha ^{\ast }\beta \right]
\eeq
and the integral
\begin{equation}
\int \frac{d^{2}z}{\pi }\exp \left[ A\left \vert z\right \vert
^{2}+Bz+Cz^{\ast }\right] =-\frac{1}{A}\exp \left[ -\frac{BC}{A}\right],
\end{equation}
the Wigner function for the three-mode entangled coherent
states can be obtained as%
\begin{equation}
W\left( \eta ,\gamma ,\delta \right) =W^{\left( 1\right) }\left( \eta
,\gamma ,\delta \right) +W^{\left( 2\right) }\left( \eta ,\gamma ,\delta
\right) \pm \left( W^{\left( 3\right) }\left( \eta ,\gamma ,\delta \right)
+W^{\left( 4\right) }\left( \eta ,\gamma ,\delta \right) \right)
\label{Super Wigner}
\end{equation}

\noindent where the different quantities occurring in Eq. (\ref{Super Wigner}) are given by
\begin{eqnarray}
W^{\left( 1\right) }\left( \eta ,\gamma \right) &=&\Theta\left \langle -z_{1},-z_{2},-z_{3}\right \vert \left(
a_{1}^{\dagger r}a_{2}^{\dagger s}a_{3}^{\dagger t}\right) \left \vert
\alpha,\alpha,\alpha\right \rangle \left \langle \alpha
,\alpha,\alpha\right \vert \left(
a_{1}^{r}a_{2}^{s}a_{3}^{t}\right) \left \vert z_{1},z_{2},z_{3}\right
\rangle \\ \nonumber
&\times& \exp \left[ 2\left( \eta z_{1}^{\ast }-\eta ^{\ast }z_{1}\right)
\right] \exp \left[ 2\left( \gamma z_{2}^{\ast }-\gamma ^{\ast }z_{2}\right)
\right] \exp \left[ 2\left( \delta z_{3}^{\ast }-\delta ^{\ast }z_{3}\right)
\right] \\ \nonumber
W^{\left( 2\right) }\left( \eta ,\gamma \right) &=&\Theta\left \langle -z_{1},-z_{2},-z_{3}\right \vert \left(
a_{1}^{\dagger r}a_{2}^{\dagger s}a_{3}^{\dagger t}\right) \left \vert
-\alpha,-\alpha,-\alpha\right \rangle\left \langle -\alpha
,-\alpha,-\alpha\right \vert \left(
a_{1}^{r}a_{2}^{s}a_{3}^{t}\right) \left \vert z_{1},z_{2},z_{3}\right
\rangle \\ \nonumber
&\times& \exp \left[ 2\left( \eta z_{1}^{\ast }-\eta ^{\ast }z_{1}\right)
\right] \exp \left[ 2\left( \gamma z_{2}^{\ast }-\gamma ^{\ast }z_{2}\right)
\right] \exp \left[ 2\left( \delta z_{3}^{\ast }-\delta ^{\ast }z_{3}\right)
\right] \\ \nonumber
W^{\left( 3\right) }\left( \eta ,\gamma \right) &=&\Theta\exp \left( i\varphi \right)\left \langle -z_{1},-z_{2},-z_{3}\right \vert \left(
a_{1}^{\dagger r}a_{2}^{\dagger s}a_{3}^{\dagger t}\right) \left \vert
-\alpha,-\alpha,-\alpha\right \rangle \left \langle \alpha
,\alpha,\alpha\right \vert \left(
a_{1}^{r}a_{2}^{s}a_{3}^{t}\right) \left \vert z_{1},z_{2},z_{3}\right
\rangle \\ \nonumber
&\times& \exp \left[ 2\left( \eta z_{1}^{\ast }-\eta ^{\ast }z_{1}\right)
\right] \exp \left[ 2\left( \gamma z_{2}^{\ast }-\gamma ^{\ast }z_{2}\right)
\right] \exp \left[ 2\left( \delta z_{3}^{\ast }-\delta ^{\ast }z_{3}\right)
\right] \\ \nonumber
W^{\left( 4\right) }\left( \eta ,\gamma \right) &=&\Theta\exp \left( -i\varphi \right)\left \langle -z_{1},-z_{2},-z_{3}\right \vert \left(
a_{1}^{\dagger r}a_{2}^{\dagger s}a_{3}^{\dagger t}\right) \left \vert
\alpha,\alpha,\alpha\right \rangle \left \langle -\alpha
,-\alpha,-\alpha\right \vert \left(
a_{1}^{r}a_{2}^{s}a_{3}^{t}\right) \left \vert z_{1},z_{2},z_{3}\right
\rangle \\  \nonumber
&\times& \exp \left[ 2\left( \eta z_{1}^{\ast }-\eta ^{\ast }z_{1}\right)
\right] \exp \left[ 2\left( \gamma z_{2}^{\ast }-\gamma ^{\ast }z_{2}\right)
\right] \exp \left[ 2\left( \delta z_{3}^{\ast }-\delta ^{\ast }z_{3}\right)
\right] \nonumber
\end{eqnarray}
with
\beq
\Theta=\left( \mathcal{N}
_{\varphi }^{r,s,t}\right) ^{2} \exp \left[
2\left( \left \vert \eta \right \vert ^{2}+\left \vert \gamma \right \vert
^{2}+\left \vert \delta \right \vert ^{2}\right) \right]  \int \frac{d^{2}z_{1}}{\pi ^{2}}\frac{d^{2}z_{2}}{\pi ^{2}}\frac{
d^{2}z_{3}}{\pi ^{2}}
\eeq
Substituting (\ref{vect-PA}) into Eq. (\ref{density OPA}) and inserting
the completeness relation of three-mode coherent states and making use of the integral formulae
\begin{eqnarray}
L_{m}\left( \zeta \xi \right) &=&\frac{e^{\zeta \xi }}{m!}\int \frac{d^{2}z}{%
\pi }z^{m}z^{\ast m}\exp \left[ -\left \vert z\right \vert ^{2}+\zeta z-\xi
z^{\ast }\right]
\end{eqnarray}

we obtain the following results
\begin{equation}
W\left( \eta ,\gamma ,\delta \right) =W^{\left( 1\right) }\left(
\eta ,\gamma ,\delta \right) +W^{\left( 2\right) }\left( \eta ,\gamma
,\delta \right) \pm \left( W^{\left( 3\right) }\left( \eta ,\gamma ,\delta
\right) +W^{\left( 4\right) }\left( \eta ,\gamma ,\delta \right) \right)
\label{Wigner exp}
\end{equation}
where
\begin{eqnarray*}
W^{\left( 1\right) }\left( \eta ,\gamma ,\delta \right) &=&\left( -1\right)
^{r+s+t}\frac{r!s!t!\left( \mathcal{N}_{\varphi }^{r,s,t}\right) ^{2}}{\pi
^{3}}\exp \left( -2\left \vert \alpha-\eta \right \vert ^{2}\right)
\exp \left( -2\left \vert \alpha-\gamma \right \vert ^{2}\right) \\
&\times& \exp \left( -2\left \vert \alpha-\delta \right \vert
^{2}\right) L_{r}\left( \left \vert \alpha-2\eta \right \vert
^{2}\right) L_{s}\left( \left \vert \alpha-2\gamma \right \vert
^{2}\right) L_{t}\left( \left \vert \alpha-2\delta \right \vert
^{2}\right)
\end{eqnarray*}
\begin{eqnarray*}
W^{\left( 2\right) }\left( \eta ,\gamma ,\delta \right) &=&\left( -1\right)
^{r+s+t}\frac{r!s!t!\left( \mathcal{N}_{\varphi }^{r,s,t}\right) ^{2}}{\pi
^{3}}\exp \left( -2\left \vert \alpha+\eta \right \vert ^{2}\right)
\exp \left( -2\left \vert \alpha+\gamma \right \vert ^{2}\right) \\
&\times& \exp \left( -2\left \vert \alpha+\delta \right \vert
^{2}\right) L_{r}\left( \left \vert \alpha+2\eta \right \vert
^{2}\right) L_{s}\left( \left \vert \alpha+2\gamma \right \vert
^{2}\right) L_{t}\left( \left \vert \alpha+2\delta \right \vert
^{2}\right)
\end{eqnarray*}
\begin{eqnarray*}
W^{\left( 3\right) }\left( \eta ,\gamma ,\delta \right) &=&\left( -1\right)
^{r+s+t}\frac{r!s!t!\left( \mathcal{N}_{\varphi }^{r,s,t}\right) ^{2}}{\pi
^{3}}\exp \left( i\varphi \right) \exp \left( -2\left \vert \eta \right
\vert ^{2}-2\eta \alpha^{\ast }+2\alpha\allowbreak \eta ^{\ast
}\right) \\
&\times & \exp \left( -2\left \vert \gamma \right \vert ^{2}-2\gamma \alpha
^{\ast }+2\alpha _{2}\allowbreak \gamma ^{\ast }\right) \exp \left(
-2\left \vert \delta \right \vert ^{2}-2\delta \alpha^{\ast }+2\alpha
_{3}\allowbreak \delta ^{\ast }\right) \\
&\times & L_{r}\left( -\left( \alpha^{\ast }-2\eta ^{\ast }\right)
\left( \alpha+2\eta \right) \right) L_{s}\left( -\left( \alpha
^{\ast }-2\gamma ^{\ast }\right) \left( \alpha+2\gamma \right)
\right) L_{t}\left( -\left( \alpha^{\ast }-2\delta ^{\ast }\right)
\left( \alpha+2\delta \right) \right)
\end{eqnarray*}%
\begin{eqnarray*}
W^{\left( 4\right) }\left( \eta ,\gamma ,\delta \right) &=&\left( -1\right)
^{r+s+t}\frac{r!s!t!\left( \mathcal{N}_{\varphi }^{r,s,t}\right) ^{2}}{\pi
^{3}}\exp \left( -i\varphi \right) \exp \left( -2\left \vert \eta \right
\vert ^{2}+2\eta \alpha^{\ast }-2\alpha\allowbreak \eta ^{\ast
}\right) \\
&\times& \exp \left( -2\left \vert \gamma \right \vert ^{2}+2\gamma \alpha
^{\ast }-2\alpha\allowbreak \gamma ^{\ast }\right) \exp \left(
-2\left \vert \delta \right \vert ^{2}+2\delta \alpha^{\ast }-2\alpha
\allowbreak \delta ^{\ast }\right) \\
&\times & L_{r}\left( -\left( \alpha^{\ast }+2\eta ^{\ast }\right)
\left( \alpha-2\eta \right) \right) L_{s}\left( -\left( \alpha
^{\ast }+2\gamma ^{\ast }\right) \left( \alpha-2\gamma \right)
\right) L_{t}\left( -\left( \alpha^{\ast }-2\delta ^{\ast }\right)
\left( \alpha+2\delta \right) \right)
\end{eqnarray*}%
In what follows we shall use the expression (\ref{Wigner exp}) to investigate
the nonclassical and non-Gaussian behaviours in (\ref{vect-PA}). In particular, the negativity of the Winger function will indicate when (\ref{vect-PA}) is non-classical.

\begin{figure}[!h]
\begin{center}
$
\begin{array}{cc}
\includegraphics[width=18cm]{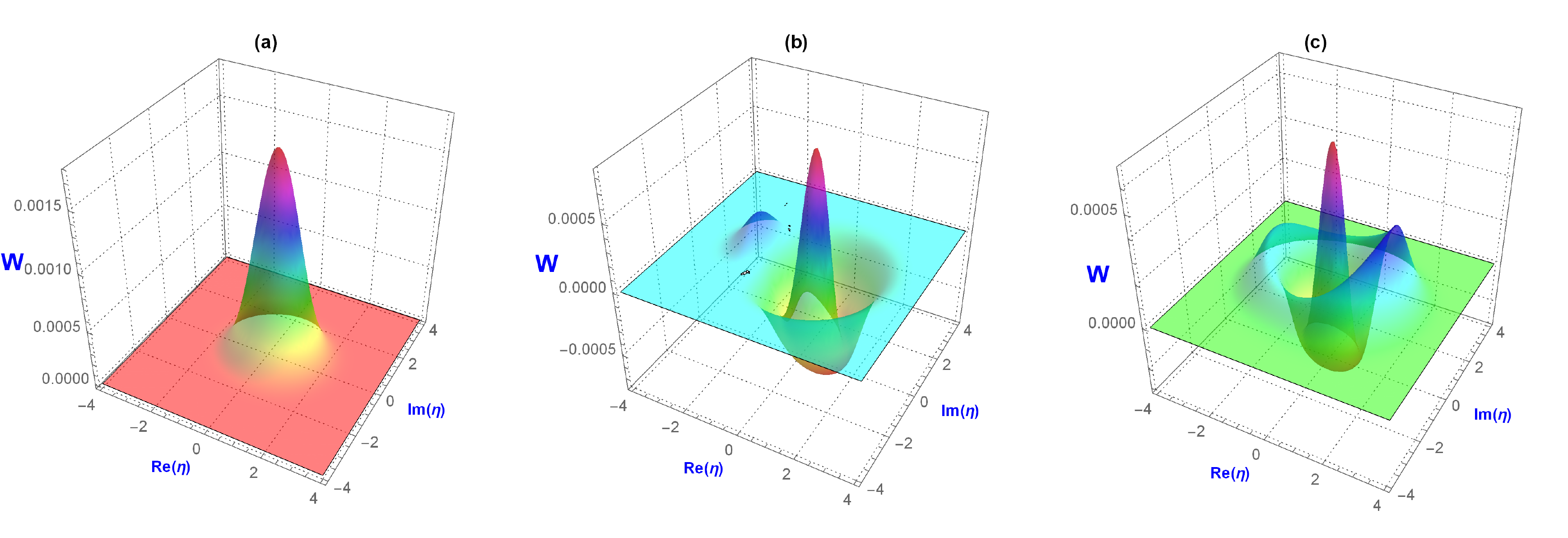}&
\end{array}
$
\end{center}
\caption{WF of photon-added entangled GHZ coherent states \eqref{Super Wigner} for even mode $\varphi =0$ versus the exchange values of $r$, $s$ and $t$; (a) $(r,s,t)=(0,0,0)$, (b) $(r,s,t)=(1,2,1)$, (c) $(r,s,t)=(2,2,2)$ with $\alpha=0.3$ and $\eta=\frac{1}{\sqrt{2}}\left( x+i y\right) $, $\gamma=\delta=1$. }
\label{fig1}
\end{figure}
\begin{figure}[!h]
\begin{center}
$
\begin{array}{cc}
\includegraphics[width=18cm]{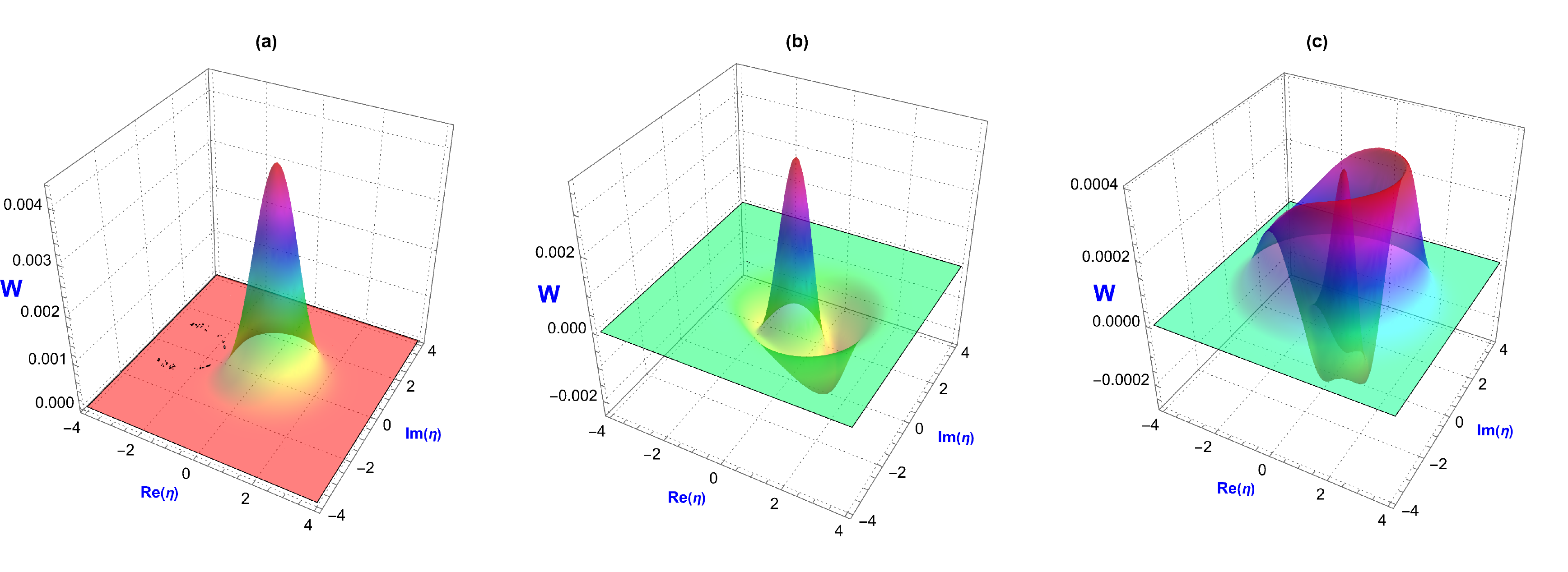}&
\end{array}
$
\end{center}
\caption{WF of photon-added entangled GHZ coherent states \eqref{Super Wigner} for even mode $\varphi =\pi$ versus the exchange values of $r$, $s$ and $t$; (a) $(r,s,t)=(0,0,0)$, (b) $(r,s,t)=(1,2,1)$, (c) $(r,s,t)=(2,2,2)$ with $\alpha=0.3$ and $\eta=\frac{1}{\sqrt{2}}\left( x+i y\right) $, $\gamma=\delta=1$.}
\label{fig2}
\end{figure}

\begin{figure}[!h]
  \centering
  \includegraphics[width=14cm]{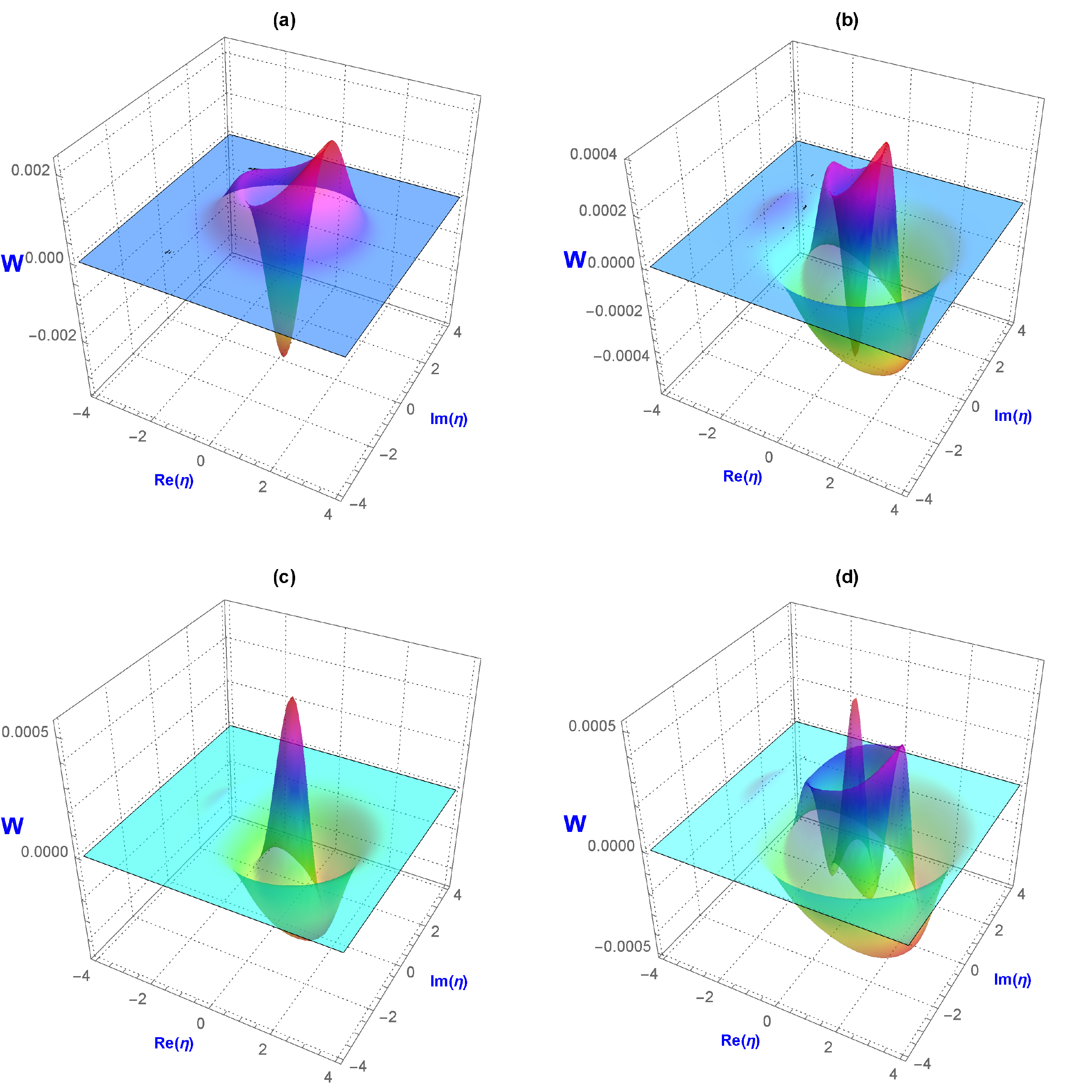}
  \caption{WF of photon-added entangled GHZ coherent states \eqref{Super Wigner} for even mode $\varphi =0$ versus the exchange values of $r$, $s$ and $t$; (a) $(r,s,t)=(1,1,0)$, (b) $(r,s,t)=(2,2,0)$, (c) $(r,s,t)=(1,2,0)$, (d) $(r,s,t)=(3,2,0)$ with $\alpha=0.3$ and $\eta=\frac{1}{\sqrt{2}}\left( x+i y\right) $, $\gamma=\delta=1$.}\label{fig3}
\end{figure}
\begin{figure}[!h]
\begin{center}
$%
\begin{array}{cc}
\includegraphics[width=14cm]{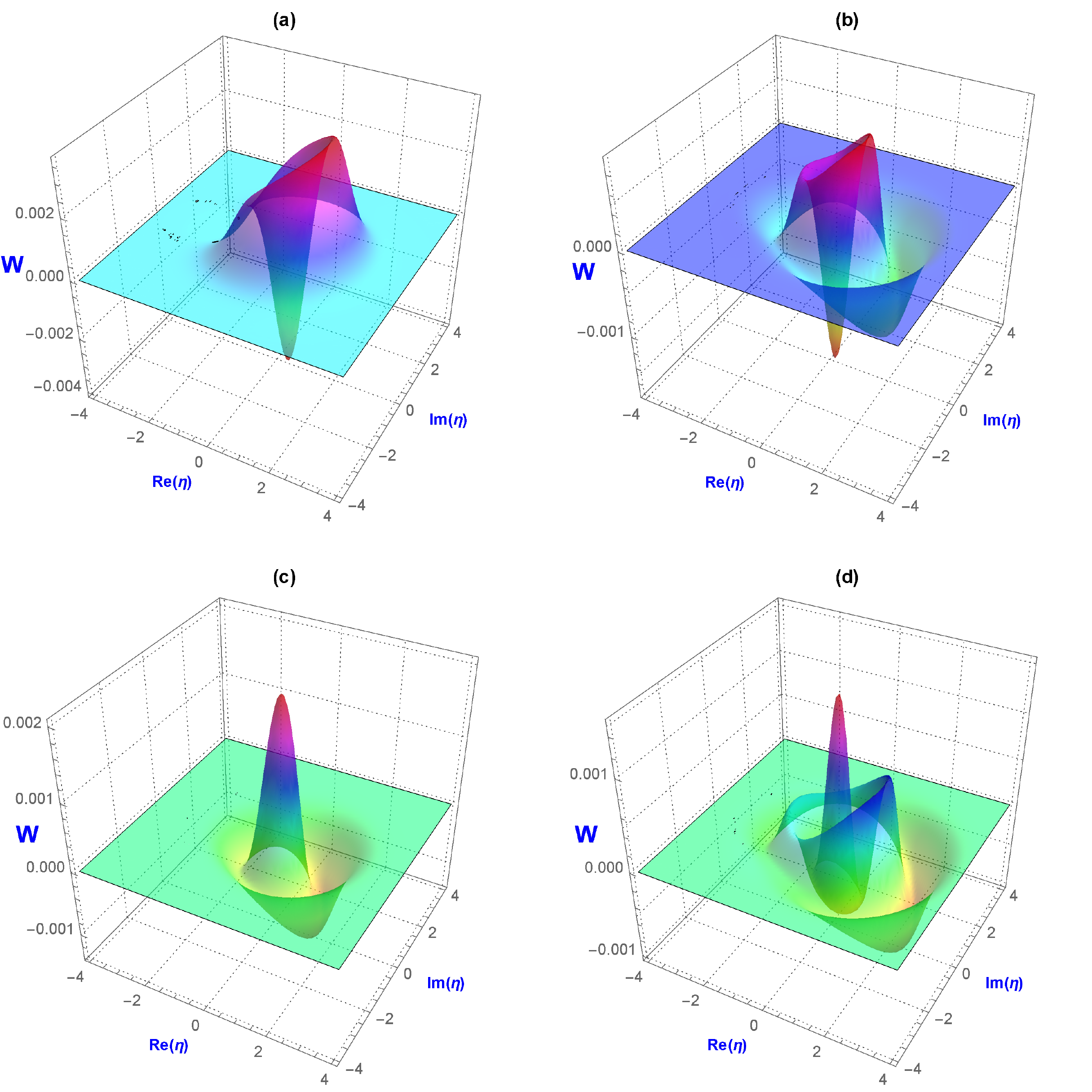} &
\end{array}%
$%
\end{center}
\caption{WF of photon-added entangled GHZ coherent states \eqref{Super Wigner} for even mode $\varphi =\pi$ versus the exchange values of $r$, $s$ and $t$; (a) $(r,s,t)=(1,1,0)$, (b) $(r,s,t)=(2,2,0)$, (c) $(r,s,t)=(1,2,0)$, (d) $(r,s,t)=(3,2,0)$ with $\alpha=0.3$ and $\eta=\frac{1}{\sqrt{2}}\left( x+\imath y\right) $, $\gamma=\delta=1$.}
\label{fig4}
\end{figure}

\begin{figure}[!h]
\begin{center}
$%
\begin{array}{cc}
\includegraphics[width=18cm]{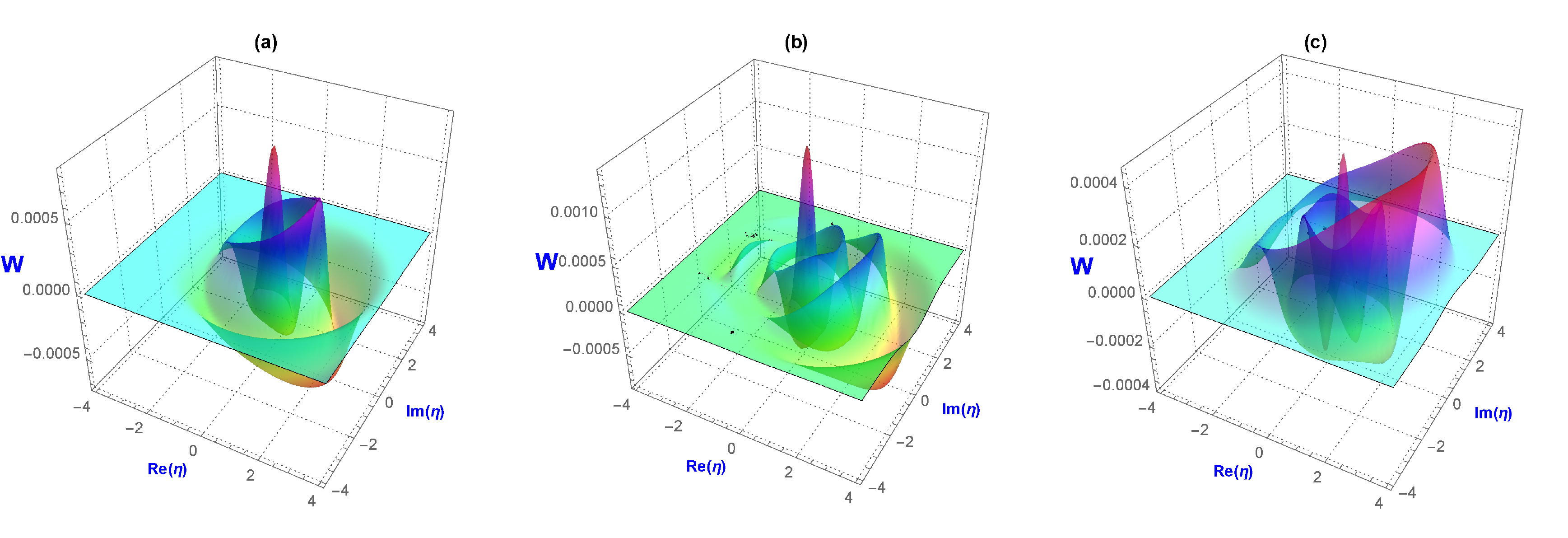} &
\end{array}%
$%
\end{center}
\caption{WF of photon-added entangled GHZ coherent states \eqref{Super Wigner} for even mode $\varphi =0$ versus the exchange values of $r$, $s$ and $t$; (a) $(r,s,t)=(3,4,5)$, (b) $(r,s,t)=(5,3,4)$, (c) $(r,s,t)=(4,5,3)$ with $\alpha=0.3$ and $\eta=\frac{1}{\sqrt{2}}\left( x+\imath y\right) $, $\gamma=\delta=1$.}
\label{fig5}
\end{figure}

\begin{figure}[!h]
\begin{center}
$%
\begin{array}{cc}
\includegraphics[width=18cm]{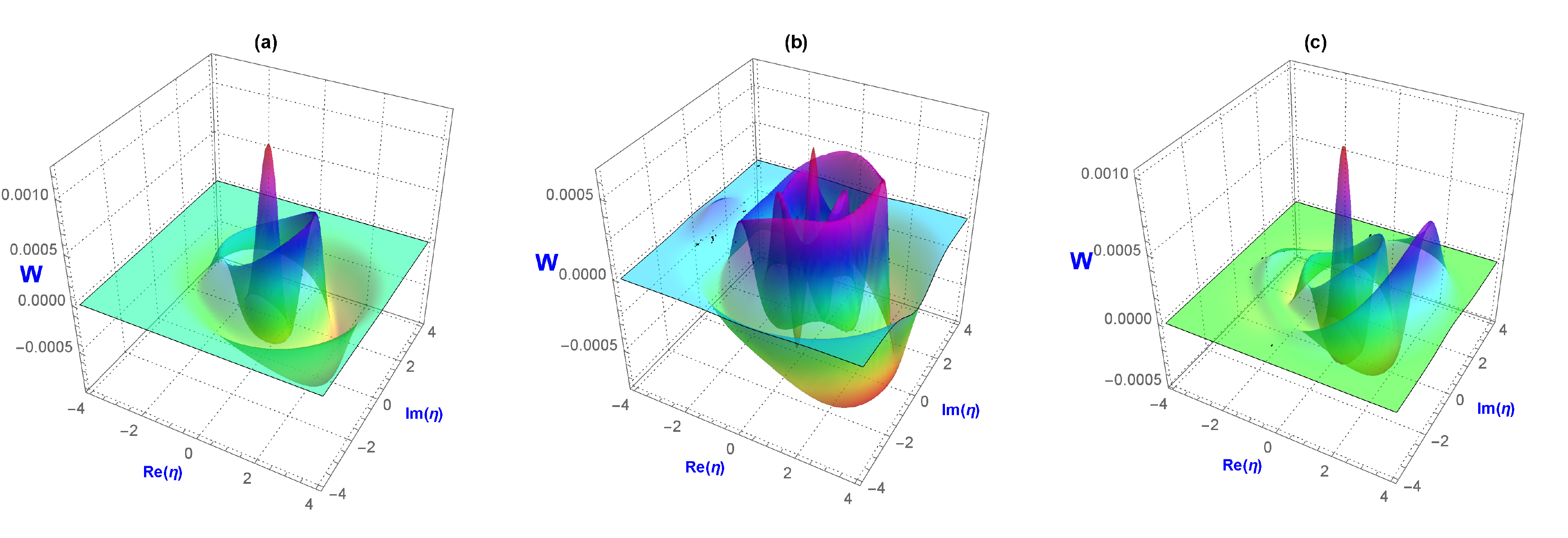} &
\end{array}%
$%
\end{center}
\caption{WF of photon-added entangled GHZ coherent states \eqref{Super Wigner} for even mode $\varphi =\pi$ versus the exchange values of $r$, $s$ and $t$; (a) $(r,s,t)=(3,4,5)$, (b) $(r,s,t)=(5,3,4)$, (c) $(r,s,t)=(4,5,3)$ with $\alpha=0.3$ and $\eta=\frac{1}{\sqrt{2}}\left( x+\imath y\right) $, $\gamma=\delta=1$.}
\label{fig6}
\end{figure}

To study the behavior of the Wigner function of the eq.(\ref{vect-PA}), we plot the variation of Wigner function in terms of the excitation photon number $r$, $s$ and $t$. When there is no photon excitation $(r=s=t=0)$, the Wigner function of state $\left \vert \psi _{0}^{r,s,t}\right \rangle $ exhibits a single upward peak at the center position and has Gaussian shape see Fig.\ref{fig1}(a), which indicates the non-classicality of the state. In addition, when only one photon is added (for instance $r=1,s=t=0$), the Wigner function can take negative values. Generally, the results show that the Wigner function gets the negative values in some regions of real and imaginary parts of $\eta$ in Fig. It confirms that the PAGHZECS is non-Gaussian and nonclassical state. Furthermore, some interesting characters of valleys are presented with increasing number of $r,s$ and $t$, which indicate that the state $\left \vert \psi _{0}^{r,s,t}\right \rangle $ is an entangled state.

The influence of the photon added number $(r,s,t)$ on Wigner function of the state $\left \vert \psi _{\pi }^{r,s,t}\right \rangle $ reported in Figs.\ref{fig2}(a), \ref{fig2}(b) and \ref{fig2}(c) is similar to
that of $\left \vert \psi _{0}^{r,s,t}\right \rangle$. Indeed, it is clear that the Wigner function has negative parts with increasing values of $r,s$ and $t$. Non-classical effects are sensitive to the excitation photon
number $r,s$ and $t$. However, we find that the negative parts are not important in comparison with the case $\left \vert \psi _{0}^{r,s,t}\right \rangle$ reported in Figs.\ref{fig1}(b) and \ref{fig1}(c).

Hence, from Figs.\ref{fig1} and Figs.\ref{fig2}  it is easy to see that, the non-classicality of the state, under consideration. For any value of $r,s$ and $t$, the non-gaussian WF of $W\left( \eta ,\gamma,\delta \right) $ always exhibits negative values in the phase space, which is another indicator of the non-classicality of the states $\left\vert \psi _{\varphi }^{r,s,t}\right \rangle $.

In Figs.\ref{fig3} and Figs.\ref{fig4}, we plot the dependence of the Winger function $W$ as a
function of both real and imaginary parts of $\eta $ with $\gamma=\delta=1$, $\varphi =0$ for Figs.\ref{fig3} and $\varphi =\pi$ for Figs.\ref{fig4}, for various values of $r,s,t$. The result shows that the Winger function of the photon added GHZ entangled coherent states takes negative values in some regions of the phase space. Therefore, we conclude that the photon added GHZ entangled coherent states in a non-classical and non-Gaussian state.

Employing Eq. (\ref{Wigner exp}), the WFs of the photon added GHZ entangled coherent state are depicted in phase space for several different photon excitation numbers $(r,s,t)$ in Figs.\ref{fig3} and \ref{fig4}. It should be noticed that the Wigner function obtained by first adding one or two photon from an initial GHZ coherent state, as can be seen from Figs.\ref{fig3}(a) and \ref{fig3}(b), when $r=s$. The non-classical character of the photon added GHZ entangled coherent state is remarkably exhibited because the WFs have a partial negative region when the photon-adding number $r=s$ and both are nonzero. In addition,  the Wigner function of state $\left \vert \psi _{\varphi}^{r,s,t}\right \rangle $ exhibits a downward peak at the center position and has Gaussian shape see Figs.\ref{fig3}(a)and \ref{fig3}(b), which indicates the non-classicality of state. This situation is also valid for the case repated in Figs.\ref{fig4}(a)and \ref{fig4}(b). Next, we consider the situation when $r\neq s$, in particular. Figures \ref{fig3}(c), \ref{fig3}(d) and Figs.\ref{fig4}(c), \ref{fig4}(d) are plotted for $(r=1,s=2)$, and $(r=3,s=2)$ (sum is even numbers), respectively, the WF of state $\left \vert \psi _{\varphi}^{r,s,t}\right \rangle $ exhibits an upward peak at the center position and has Gaussian shape, which indicates the non-classicality of state. On the other hand, one can clearly see that the negativity of the Winger function depends on the number of photons added and on the nature of the initial state, symmetric or antisymmetric. However, the Winger function associated $\left \vert \psi _{0}^{r,s,t}\right \rangle $ exhibits more non-classicality (see Fig. \ref{fig3}). Thus, we can conclude that case $\left \vert \psi _{0}^{r,s,t}\right \rangle $ shows stronger non-classical behavior than case $\left \vert \psi _{\pi}^{r,s,t}\right \rangle $.

In the next, to see the behavior of the WF of the PAGHZECSs, we plot the three dimensional graphics with varying excitation photon number in Fig. \ref{fig5} and \ref{fig6}. Where, the numbers of photon added $r$, $s$ and $t$ switch the values $3$, $4$ and $5$ between them, respectively. We observe similar behaviours in comparison between Figs. \ref{fig5}(a), \ref{fig5}(b) and \ref{fig5}(c) for $\varphi=0$ and Figs. \ref{fig6}(a), \ref{fig6}(b) and \ref{fig6}(c) for $\varphi=\pi$.

\subsection{Sub-Poissonian photon statistics}
Here, we study the photon number statistics for the quantum states under consideration by evaluating
the Mandel parameter. This parameter is a measure of the sub-Poissonian
statistics and it is defined as the normalized variance of the photon number
distribution and for each mode as follows
\begin{equation*}
Q_{i}=\frac{\left \langle
\widehat{n}_{i}^{2}\right \rangle -\left( \left \langle \widehat{n}%
_{i}\right \rangle \right) ^{2}}{\left \langle \widehat{n}_{i}\right \rangle
}=\frac{\left \langle \left( a_{i}^{\dagger }a_{i}\right) ^{2}\right \rangle
}{\left \langle a_{i}^{\dagger }a_{i}\right \rangle }-\left \langle
a_{i}^{\dagger }a_{i}\right \rangle ,\text{ \  \ }i=1,2,3
\end{equation*}
where $\widehat{n}_{i}=a_{i}^{\dagger }a_{i} (i=1,2,3)$ are the number operators
corresponding the $i$ (the subscript $i$ relates to the $ith$ mode).

For positive values of the Mandel parameter $Q$ we have
super-Poissonian statistics (classical states), zero value $\left(
Q=0\right) $ corresponds to coherent state $\left
\vert \alpha \right \rangle $ and negative values $\left(
Q<0\right) $ represents, Poissonian and
sub-Poissonian photon statistics which reflect a non-classical
character of the states.

To evaluate Mandel's $Q$ factors, we first compute the average photon number of each mode of the photon added GHZ entangled coherent state
\begin{eqnarray}
\left\langle a_{1}^{\dagger }a_{1}\right\rangle =\frac{\mathcal{N}_{\varphi
}^{r+1,s,t}}{\mathcal{N}_{\varphi }^{r,s,t}}-1,\text{ }\left\langle
a_{2}^{\dagger }a_{2}\right\rangle =\frac{\mathcal{N}_{\varphi }^{r,s+1,t}}{%
\mathcal{N}_{\varphi }^{r,s,t}}-1,\text{ }\left\langle a_{3}^{\dagger
}a_{3}\right\rangle =\frac{\mathcal{N}_{\varphi }^{r,s,t+1}}{\mathcal{N}%
_{\varphi }^{r,s,t}}-1
\end{eqnarray}
where ${\mathcal{N}_{\varphi }^{r,s,t}}$ is given by (\ref{Normaliz PA}).
The expectation values of the operators $a_{i}^{\dagger 2}a_{i}^{2}$ are
\begin{eqnarray}\label{Numberph}
\left \langle a_{1}^{\dagger 2}a_{1}^{2}\right \rangle &=&\frac{\mathcal{N}_{\varphi }^{r+2,s,t}-4\mathcal{N}_{\varphi }^{r+1,s,t}}{\mathcal{N}
_{\varphi }^{r,s,t}}+2 \\ \nonumber
\left \langle a_{2}^{\dagger 2}a_{2}^{2}\right \rangle &=&\frac{\mathcal{N}_{\varphi }^{r,s+2,t}-4\mathcal{N}_{\varphi }^{r,s+1,t}}{\mathcal{N}_{\varphi }^{r,s,t}}+2 \\ \nonumber
\left\langle a_{3}^{\dagger 2}a_{3}^{2}\right \rangle &=&\frac{\mathcal{N}_{\varphi
}^{r,s,t+2}-4\mathcal{N}_{\varphi }^{r,s,t+1}}{\mathcal{N}_{\varphi }^{r,s,t}
}+2
\end{eqnarray}

Thus, one obtains the Mandel's $Q_{1}$, $Q_{2}$ and $Q_{3}$ parameters of the
PAGHZECS, they are given by
\begin{eqnarray}
Q_{1} &=&\frac{\mathcal{N}_{\varphi }^{r+2,s,t}-4\mathcal{N}%
_{\varphi }^{r+1,s,t}+2\mathcal{N}_{\varphi }^{r,s,t}}{\mathcal{N}_{\varphi
}^{r+1,s,t}-\mathcal{N}_{\varphi }^{r,s,t}}-\frac{\mathcal{N}_{\varphi
}^{r+1,s,t}}{\mathcal{N}_{\varphi }^{r,s,t}}+1  \label{Qmandel} \\
Q_{2} &=&\frac{\mathcal{N}_{\varphi }^{r,s+2,t}-4\mathcal{N}%
_{\varphi }^{r,s+1,t}+2\mathcal{N}_{\varphi }^{r,s,t}}{\mathcal{N}_{\varphi
}^{r,s+1,t}-\mathcal{N}_{\varphi }^{r,s,t}}-\frac{\mathcal{N}_{\varphi
}^{r,s+1,t}}{\mathcal{N}_{\varphi }^{r,s,t}}+1  \notag \\
Q_{3} &=&\frac{\mathcal{N}_{\varphi }^{r,s,t+2}-4\mathcal{N}%
_{\varphi }^{r,s,t+1}+2\mathcal{N}_{\varphi }^{r,s,t}}{\mathcal{N}_{\varphi
}^{r,s,t+1}-\mathcal{N}_{\varphi }^{r,s,t}}-\frac{\mathcal{N}_{\varphi
}^{r,s,t+1}}{\mathcal{N}_{\varphi }^{r,s,t}}+1  \notag
\end{eqnarray}

\begin{figure}[!h]
\begin{center}
\includegraphics[width=16cm]{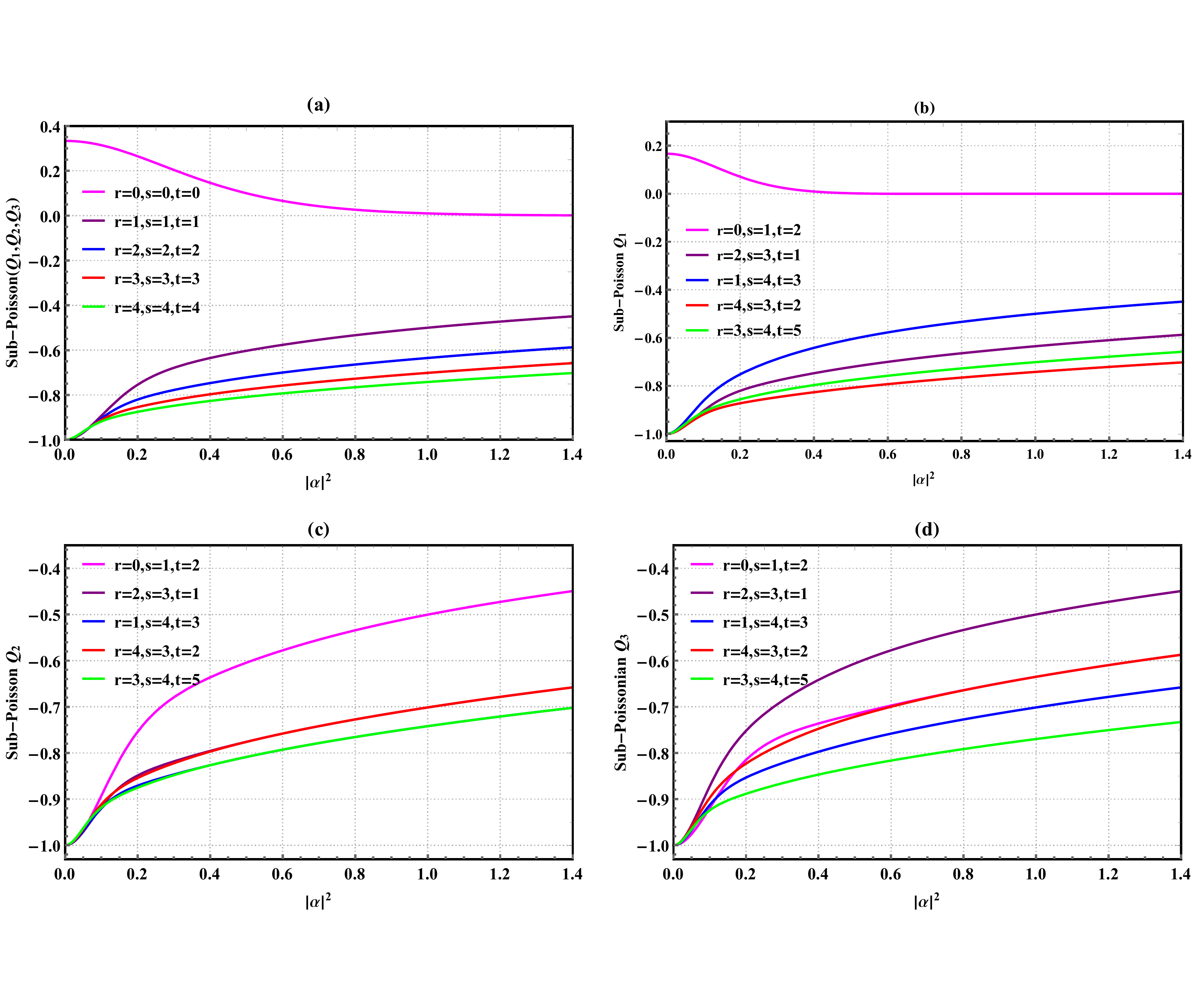}
\end{center}
\caption{Mandel parameter Q as a function of the coherent state
parameter $\left \vert \alpha \right \vert ^{2}$ for even mode
$\protect \varphi =0$ and various values of $(r,s,t)$ of photon
added;{\color{black} (a) Mandel's $Q$ factors of $Q_{1}=Q_{2}=Q_{3}$ with $(r=s=t)$, (b) Mandel's factor of $Q_{1}$ with $(r\neq s \neq t)$, (c) Mandel's factor of $Q_{2}$ with $(r\neq s \neq t)$, (d) Mandel's factor of $Q_{3}$ with $(r\neq s \neq t)$}.  } \label{fig7}
\end{figure}
\begin{figure}[!h]
\begin{center}
\includegraphics[width=16cm]{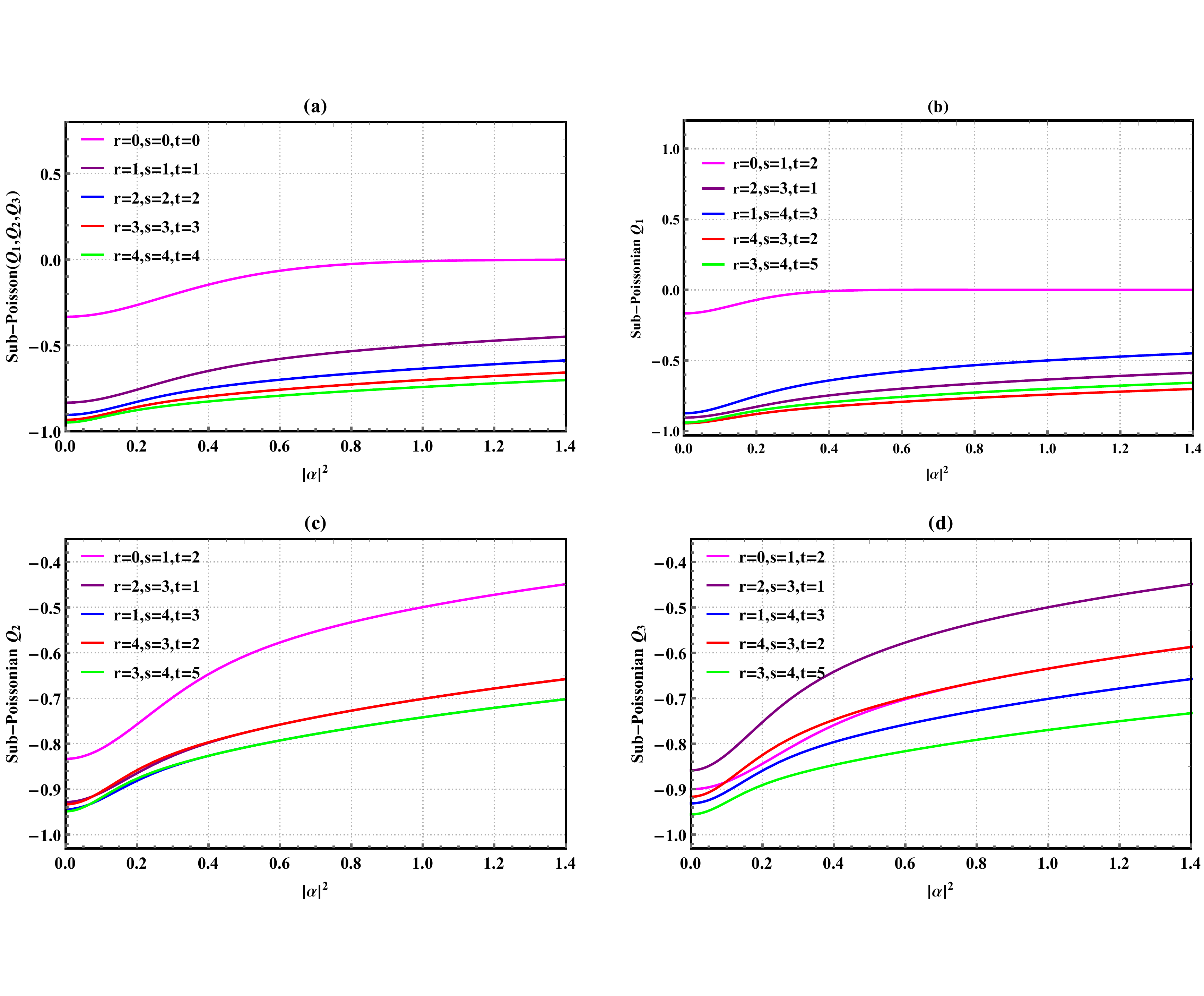}
\end{center}
\caption{Mandel parameter Q as a function of the coherent state
parameter $\left \vert \alpha \right \vert ^{2}$ for even mode
$\protect \varphi =\pi$ and various values of $(r,s,t)$ of photon
added;{\color{black} (a) Mandel's $Q$ factors of $Q_{1}=Q_{2}=Q_{3}$ with $(r=s=t)$, (b) Mandel's factor of $Q_{1}$ with $(r\neq s \neq t)$, (c) Mandel's factor of $Q_{2}$ with $(r\neq s \neq t)$, (d) Mandel's factor of $Q_{3}$ with $(r\neq s \neq t)$ }.} \label{fig8}
\end{figure}

The Mandel parameter for each mode of the symmetric state $\varphi =0$(even) and
antisymmetric state $\varphi =\pi $(odd) \textquotedblleft is plotted versus $\left \vert \alpha \right \vert ^{2}$
for different values of photon-added modes, $r$, $s$ and
$t$. The results are reported in Fig. \ref{fig7} and \ref{fig8} with $\varphi =0$ and
$\varphi =\pi$ respectively. \

In Fig. \ref{fig7}, we plot Mandel's parameter $Q_{i}^{\left \vert \psi _{\varphi }\right \rangle}$ for $\varphi=0$ versus $\left
\vert \alpha \right \vert ^{2}$ associated of each mode such that $Q_{1}$ correspond to
mode $1$, $Q_{2}$ corresponds to mode $2$ and $Q_{3}$ corresponds to mode $3$, have been presenteds them and compared
as a function of the parameter $\left \vert \alpha \right \vert
^{2}$, and for various values of
photon added $r$, $s$ and $t$ with $\left( \varphi =0\right) $ (even
 state). As Fig. \ref{fig7} shows, all modes of state $\left \vert \psi _{\varphi}^{r,s,t}\right \rangle $ represent fully
sub-Poissonian statistics for any values of $(r,s,t)$ and $\left
\vert \alpha \right \vert ^{2}$. However different statistics
emerges due to different choice of the modes. For instance, based on
Fig. \ref{fig7}, for fixed small values of $\left \vert
\alpha \right \vert ^{2}$, the measure of non-classicality decreases,
when number of photons added $(r,s,t)$ is enhanced. Specifically,
Figs. \ref{fig7} reveals that the measure of non-classicality of
the modes $Q_{2}$ and $Q_{3}$ become larger
than the mode $Q_{1}$. In other words, different
statistics may be obtained depending are numbers of photon-added
in even GHZ coherent states. In addition, Figs. \ref{fig7}(a) and \ref{fig7}(b)for $(r=s=t=0)$ and $(r=0,s=1,t=2)$ it is easy to see that the initial super-Poissonian statistics $(Q>0)$ for $|\alpha|^{2}<0.4$ is rapidly transformed into the coherent states $(Q=0)$ with the increasing values of $|\alpha|^{2}>0.4$. Furthermore, for $(r\neq 0,s\neq 0,t\neq 0)$  in Figs. \ref{fig7}(a) and \ref{fig7}(b) we have $(Q<0)$ for sub-Poissonian statistics, and same behaviour viewed in Figs. \ref{fig7}(c) and \ref{fig7}(d)for any values of $r$, $s$ and $t$. Moreover, in the mode odd $\varphi=\pi$ it can be formed from Fig. \ref{fig8} for $(r=s=t=0)$ and $(r=0,s=1,t=2)$ in Figs. \ref{fig8}(a) and \ref{fig8}(b) it is easy to see that the initial sub-Poissonian statistics $(Q<0)$ for $|\alpha|^{2}<1$ is slowly transformed into the coherent state $(Q=0)$ for $|\alpha|^{2}>1$. But, we have the same behaviour observed in Fig. \ref{fig7} also obtained in Fig. \ref{fig8} in each mode.
Clearly, the Mandel parameter of the each mode even/odd $\left( \varphi =0\right)$ in Fig. \ref{fig7} and Fig. \ref{fig8} show the same behaviors. It can be observed a strong non-classical property in each mode even/odd for all values of $(r\neq 0,s\neq 0,t\neq 0)$, the non-classicality measure increases when the $r$, $s$ and $t$ increases for all values of $\left \vert \alpha \right \vert ^{2}$. While on each mode of the even/odd three-mode photon-added GHZ entangled coherent states becomes
non-classical for all consideration values of photon-additions $(r,s,t)$
and all values of $\left \vert \alpha \right \vert ^{2}$.
In the case of the mode odd $\left( \varphi =\pi \right) $ see Fig.
\ref{fig8} three-mode photon-added GHZ entangled coherent state of the each mode
coherent states, the non-classicality has absolutely exhibits the
higher-order in all the considered situations but it is important in two cases when
$(r=s=t)$ and $(r\neq s\neq t)$. We can conclude that adding photon
in three modes GHZ entangled coherent state has a great effect on increasing
the non-classicality feature of the each mode of the odd state and
approaches to the value zero when the parameter $\left \vert
\alpha \right \vert ^{2}$ is large.

\bigskip

\subsection{Second order correlation function}
{\color{black} The analytical expression of correlation functions at any order, taking into
consideration the non-unit quantum efficiency of the detection scheme,
utilizing only values that can be obtained experimentally by direct
detection We also illustrate that high-order correlations serve as a
valuable tool for determining the nature of the state and show that as
correlation order increases, the distinctions between classical and quantum
states become increasingly clear.

The correlation functions $g_{\widehat{n}}^{jk}$ are usually defined in terms of the
normally-ordered creation and annihilation operators \cite{Allevi16}.
\begin{equation}
g_{\widehat{n}}^{jk}=\frac{\left \langle a_{1}^{\dagger
j}a_{1}^{j}a_{2}^{\dagger k}a_{2}^{k}\right \rangle }{\left \langle
a_{1}^{\dagger }a_{1}\right \rangle ^{j}\left \langle a_{2}^{\dagger
}a_{2}\right \rangle ^{k}}
\end{equation}
where $a_{k}$ is the operator of the mode k-th and $\widehat{n}_{k}=a_{k}^{\dagger }a_{k}$.
Thus they introduce the second-order correlation function $g^{(2)}(0)$ \cite%
{Glauber63, Loudon83},which leads to better understanding of the
non-classical behavior of the quantum states \cite{Penna38}.

In this section, we generalized extension for the three-mode correlation
function is given by \cite{Mujahid, Khan, Abdisa}

\begin{equation}
g_{123}^{\left( 3\right) }\left( 0\right) =\frac{\left \langle
a_{1}^{\dagger }a_{1}a_{2}^{\dagger }a_{2}a_{3}^{\dagger }a_{3}\right
\rangle }{\left \langle a_{1}^{\dagger }a_{1}\right \rangle \left \langle
a_{2}^{\dagger }a_{2}\right \rangle \left \langle a_{3}^{\dagger
}a_{3}\right \rangle },  \label{Correlation}
\end{equation}
where $i, j, k = 1, 2, 3$ and $i\neq j\neq k$.}
\noindent The expectation value of $\left \langle a_{1}^{\dagger }a_{1}a_{2}^{\dagger
}a_{2}a_{3}^{\dagger }a_{3}\right \rangle $ in the PAGHZECS is%
\begin{equation}
\left \langle a_{1}^{\dagger }a_{1}a_{2}^{\dagger }a_{2}a_{3}^{\dagger
}a_{3}\right \rangle =\frac{\mathcal{N}_{\varphi }^{r+1,s+1,t+1}-\mathcal{N}%
_{\varphi }^{r+1,s,t}-\mathcal{N}_{\varphi }^{r,s+1,t}-\mathcal{N}_{\varphi
}^{r,s,t+1}}{\mathcal{N}_{\varphi }^{r,s,t}}+1  \label{29}
\end{equation}

In the situation when the function $g_{123}^{\left( 3\right) }\left( 0\right) $ is positive we have the photon bunching, and $g_{123}^{\left( 3\right) }\left( 0\right)<0 $ we have the photon anti-bunching

\noindent Reporting Eqs. (\ref{Numberph}) and (\ref{29}) into Eq.
(\ref{Correlation}), we obtain the cross-correlation function as follows

\begin{equation}
g_{123}^{\left( 3\right) }\left( 0\right) =\left( \mathcal{N}_{\varphi
}^{r,s,t}\right) ^{2}\frac{\mathcal{N}_{\varphi }^{r+1,s+1,t+1}-\mathcal{N}%
_{\varphi }^{r+1,s,t}-\mathcal{N}_{\varphi }^{r,s+1,t}-\mathcal{N}_{\varphi
}^{r,s,t+1}+\mathcal{N}_{\varphi }^{r,s,t}}{\left( \mathcal{N}_{\varphi
}^{r+1,s,t}-\mathcal{N}_{\varphi }^{r,s,t}\right) \left( \mathcal{N}%
_{\varphi }^{r,s+1,t}-\mathcal{N}_{\varphi }^{r,s,t}\right) \left( \mathcal{N%
}_{\varphi }^{r,s,t+1}-\mathcal{N}_{\varphi }^{r,s,t}\right) }-1
\label{Scorrelation}
\end{equation}

\noindent and it becomes less than $1$ for a nonclassical state $(g_{abc}^{\left(
3\right) }\left( 0\right) <1)$.

\begin{figure}[th]
\begin{center}
\includegraphics[width=16cm]{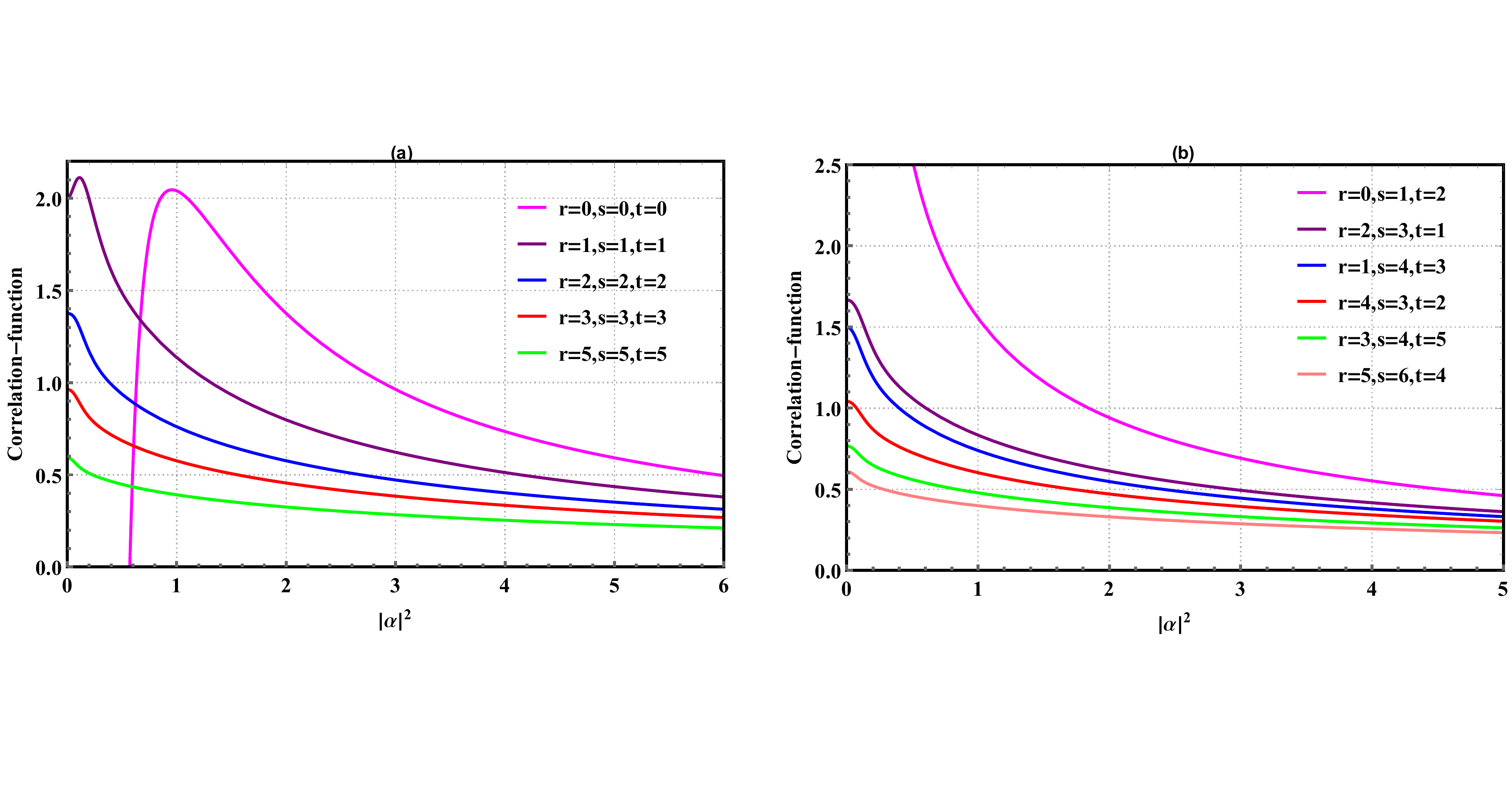}
\end{center}
\caption{Second order correlation $g_{abc}^{\left( 3\right) }\left(
0\right)$  as a function of the coherent state parameter $\left
\vert \alpha \right \vert ^{2}$ for even mode $\protect \varphi =0$
and various values of $(r,s,t)$ of photon added;{\color{black} (a) $(r=s=t)$, (b) $(r\neq s\neq t)$}.} \label{fig9}
\end{figure}

\begin{figure}[!h]
\begin{center}
\includegraphics[width=16cm]{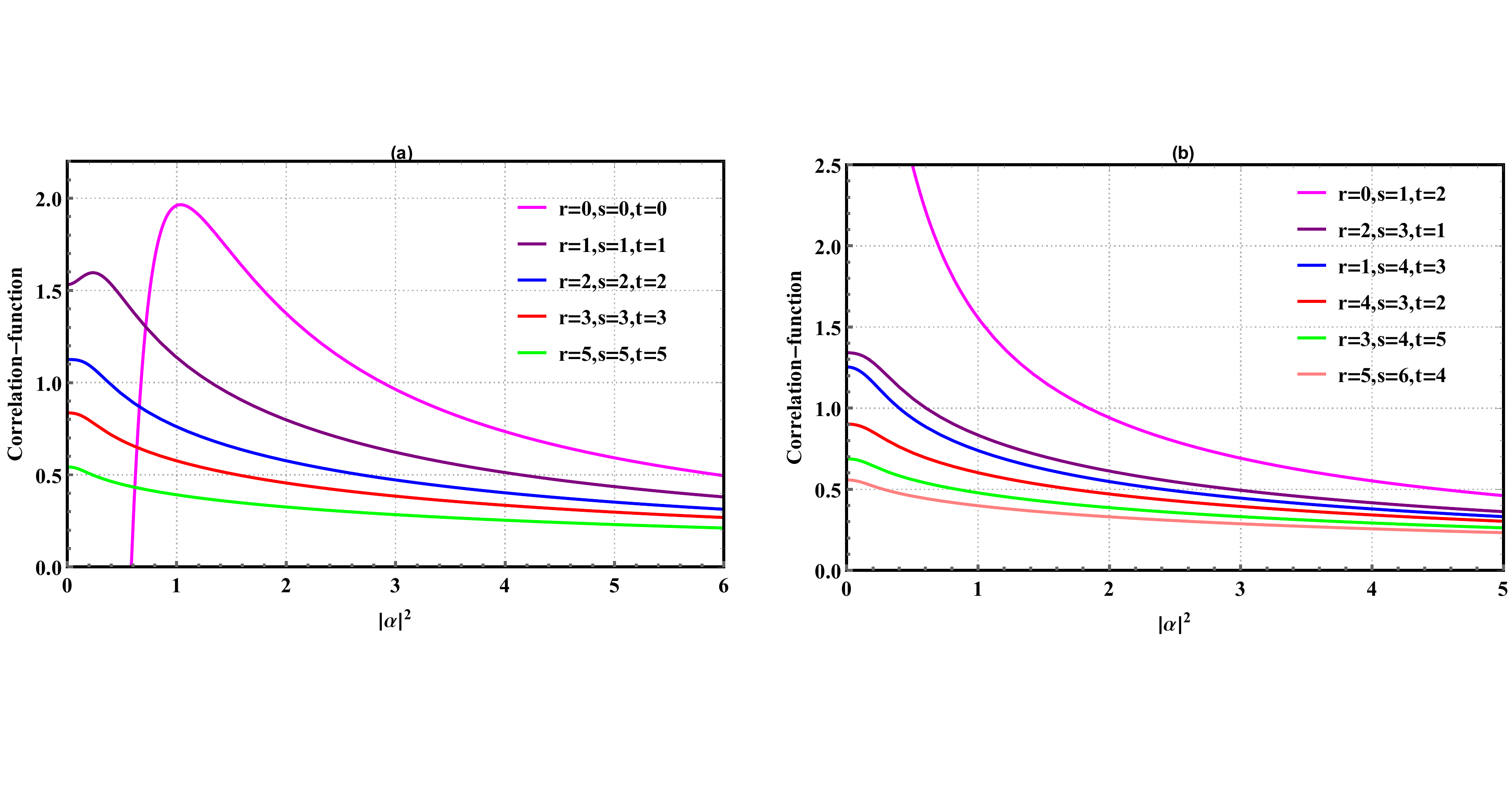}
\end{center}
\caption{Second order correlation $g_{abc}^{\left( 3\right) }\left(
0\right)$  as a function of the coherent state parameter $\left
\vert \alpha \right \vert ^{2}$ for even mode $\protect \varphi =\pi$
and various values of $(r,s,t)$ of photon added;{\color{black} (a) $(r=s=t)$, (b) $(r\neq s\neq t)$}.} \label{fig10}
\end{figure}

We plot the second-order correlation function Eq. (\ref{Scorrelation}) versus different values of
$r$, $s$ and $t$ in Figs. \ref{fig9} and \ref{fig10} with $\varphi
=0$ and $\varphi =\pi$ respectively.

\noindent We examine the second order correlation function by using Eq. (\ref{Scorrelation}). In Figs. \ref{fig9}(a), \ref{fig9}(b) and Figs. \ref{fig10}(a), \ref{fig10}(b), we plot the dependence of $g_{abc}^{\left( 3\right) }\left( 0\right) $ on $\left \vert \alpha
\right \vert ^{2}$ for several values of $\left( r,s,t\right) $,
therein over all of the regions of $\left \vert \alpha \right \vert
^{2}$, the case of $r=s=t=0$ (the solid Pink line) corresponds to
the GHZECS. The graphs of second-order correlation function in
Fig. \ref{fig9} and \ref{fig10} clearly indicates the anti-bunching
phenomenon.

\bigskip
\newpage
\section{Concluding remarks}


In this paper, we have introduced a class of state called photon-added three modes GHZ coherent state and studied their nonclassical and non-Gaussian properties based on the Wigner function. It is shown that the Wigner function of the photon-added three modes GHZ coherent state gets negative values in some regions of the phase space and depends of the number on added photons to the three modes GHZ coherent states. This shows that the photon-added three modes GHZ coherent state quantum features is affected by this effect and it becomes a nonclassical and non-Gaussian state, whereas the original GHZ coherent state is the Gaussian state. The nonclassical and non-Gaussian properties of this state occur by adding photons to the original GHZ coherent state. In addition, the obtained results show that the Mandel parameter of the photon-added three modes of the even and odd GHZ coherent state always presents negative values; this indicates that the photon-added three modes GHZ coherent state obeys sub-Poissonian statistics, characteristic of non-classicality. However, the sub-Poissonian characteristics of the three modes of these entangled states increase with increasing the photon-addition of the mode $r$, $s$ and $t$. Furthermore, the second-order correlation function does not show any non-classical feature for the even and odd three-mode photon-added GHZ entangled coherent states for all considered values of photon additions $r$, $s$ and $t$.
This is another interstice results of this work and we hope to be able to generalize this to the case of $N>3$ partite coherent states.

\end{document}